\documentclass[fleqn,10pt]{wlscirep}
\usepackage[utf8]{inputenc}
\usepackage[T1]{fontenc}
\usepackage{float} 

\title{Hydrogen Diffusion in Magnesium Using Machine Learning Potentials: a comparative study}
\author[1,+]{Andrea Angeletti}
\author[2,*,+]{Luca Leoni}
\author[3]{Dario Massa}
\author[2]{Luca Pasquini}
\author[3]{Stefanos Papanikolaou}
\author[1,2]{Cesare Franchini}
\affil[1]{University of Vienna, Vienna Doctoral School in Physics, Boltzmanngasse 5, 1090, Vienna, Austria, Europe}
\affil[2]{Department of Physics and Astronomy 'Augusto Righi', Alma Mater Studiorum - Università di Bologna, Viale Berti Pichat 6/2, 40127, Bologna, Italy, Europe}
\affil[3]{NOMATEN Centre of Excellence, National Center for Nuclear Research, 05-400, Otwock, Poland, Europe}
\affil[*]{Corresponding author: luca.leoni12@unibo.it}
\affil[+]{These authors contributed equally to this work}

\graphicspath{{./}}

\newcommand{\blue}[1]{\textcolor{black}{#1}}

\begin{abstract}
Understanding and accurately predicting hydrogen diffusion in materials is challenging due to the complex interactions between hydrogen defects and the crystal lattice. These interactions span large length and time scales, making them difficult to address with standard ab initio techniques. This work addresses this challenge by employing accelerated machine learning (ML) molecular dynamics simulations through active learning.
We conduct a comparative study of different ML-based interatomic potential schemes, including VASP, MACE, and CHGNet, utilizing various training strategies such as on-the-fly learning, pre-trained universal models, and fine-tuning.
\blue{By considering different temperatures and concentration regimes, we} obtain hydrogen diffusion coefficients 
\blue{and activation energy values}
which align \blue{remarkably} well with experimental results, underlining the efficacy and accuracy of ML-assisted methodologies in the context of diffusive dynamics. 
Particularly, our procedure significantly reduces the computational effort associated with traditional transition state calculations or ad-hoc designed interatomic potentials.
The results highlight the limitations of pre-trained universal solutions for defective materials and how they can be improved by fine-tuning. 
Specifically, fine-tuning the models on a database produced during on-the-fly training of VASP ML force\blue{-}field allows the retrieving of DFT-level accuracy at a fraction of the computational cost.
\end{abstract}
\begin{document}

\flushbottom
\maketitle
%
%
\section{Introduction}
The global imperative for sustainable and green energy solutions has intensified the search for efficient hydrogen storage materials. With the highest energy density among fuels~\cite{Yang2010656}, hydrogen represents a promising alternative to fossil sources that can be produced with zero CO$_2$ emissions powered by surplus renewable energy\cite{surplus}, through methods such as electrolysis~\cite{TARASOV202113647,Hfromrenew,Hfromrenew2}. 
The main barrier preventing a future economy based on hydrogen energy is \blue{the cost of production}, and the absence of a green, safe and efficient way to store and transport it.
Solid state hydrogen storage technologies are the most studied in this regard, being the safest, offering higher volumetric densities~\cite{safetyofsolid} than cryogenic or high-pressure gaseous alternatives~\cite{highPinefficient_1, highPinefficient_2, highPinefficient_3}.
Despite these advantages, the technology remains in its early stages and the search for materials allowing large-scale applications, like in the automotive industry~\cite{udsoe}, remains open~\cite{yartys2019, hirscher2020}.

\blue{Metal hydrides currently represent promising efficient and economical solutions~\cite{metal}.} 
Particularly, Magnesium stands out for its excellent hydrogen storage capacity~\cite{storage}, environmental friendliness, and natural abundance, displaying theoretical storage capacities as high as 7.6\% wt \cite{wt}.
However, the slow kinetics of hydrogen in magnesium-based compounds still pose a limit to possible applications.
Therefore, understanding and optimizing hydrogen diffusion pathways through theoretical modeling~\cite{dftmd, dftneb, dftreview} and experimental studies~\cite{nishimure, pasquini,Renner}  is crucial in order to improve performances of future Mg-based hydrogen storage materials.
Despite numerous efforts over the past decade, modeling hydrogen dynamics in solid-state compounds remains challenging~\cite{bccml, reinforcement,palladium,binary}. 
The low hydrogen diffusivity in magnesium requires prolonged simulation times, in the order of nanoseconds, for accurate studies using ab-initio Molecular Dynamics (MD). 
Consequently, ab-initio transition state calculations, such as the nudged elastic band (NEB) method, have emerged as the most effective approaches to reproduce and interpret the experimental data documented in the literature to date~\cite{dftneb}.
However, this technique is cumbersome and often impracticable for systems with high defect concentrations, or complex potential energy landscapes: manually describing all possible paths required by NEB may be very challenging and virtually impossible~\cite{dftneb}. 

Recently, Machine Learning accelerated MD (MLMD) has revolutionized the world of MD by making accurate simulations of large systems accessible over long time scales~\cite{Friederich2021}.
The application of such approach to study hydrogen defective systems is of high interest\cite{zrh2}, since the prediction of dynamical properties would highly expand the limited landscape offered by today\blue{'s} transition state computations.
MLMD does allow the study of multi-component system\blue{s}~\cite{vandenhaute2023machine} and can efficiently account for interaction between defects~\cite{vacancy}.
However, developing accurate interatomic potentials, especially for hydrogen-defective materials, is notoriously challenging~\cite{cluster, ml}. 
Still, the field is rapidly growing and several new approaches were proposed to explore new and complex phase spaces.
On one side, various pre-trained universal solutions~\cite{chgnet, mace} start to be available, aiming to offer a convenient and versatile way \blue{of} tackling the problem. However, as discussed in this work, while their training datasets are rich in chemical compositional space, the limited configurational sampling can significantly compromise their accuracy on previously unseen defected, metastable and transition states, leading to ungranted generalization capabilities.
On the other side, active-learning approaches based on Bayesian force fields are showing a large versatility thanks to the construction of on-the-fly databases~\cite{mlff1, mlff2}.
The error-oriented sampling of configurations allows these models to easily collect high-quality data widely spanning the configurational space, thus making them highly accurate despite their architecture, constrained compared to neural networks.
The current study aims at illustrating a systematic procedure for ML-potentials applications in diffusive dynamics conditions, which can be used to enhance the study of different
embedded defects, without departing from ab-initio accuracy. 
This procedure specifically consists in improving ML-based pre-trained models' performance via actively learned configurations, generated by on-the-fly training of the Vienna Ab Initio Simulation Package Bayesian ML Force-Field (VASP-MLFF)\cite{mlff1,mlff2}. 
\blue{We consider four different concentrations (MgH$_{0.03125}$, MgH$_{0.046875}$, MgH$_{0.0625}$ and MgH$_{0.078125}$) and} compute the hydrogen diffusion coefficient at three different temperatures \blue{(300~K, 480~K, and 673~K)}, employing a proper methodology which ensures accurate analysis of unbiased dynamical properties.
Two different Universal Interatomic Potentials (UIPs), CHGNet\cite{chgnet} and MACE\cite{mace}, were considered. 
Dynamical properties were computed for VASP-MLFF alongside the pre-trained and fine-tuned versions of the UIPs.
The comparison between the different results and experimental data showed excellent agreement, both with VASP-MLFF and fine-tuned potentials, while \blue{the} pre-trained versions fail to reach a satisfactory accuracy.
Interestingly, the fine-tuned potentials outperform VASP-MLFF by correctly predicting the temperature dependence of the diffusion coefficient.
\section{Material and Methods}
\subsection{MLFF-MD}
\begin{figure}[b!]
\centering
\includegraphics[width=0.6\linewidth]{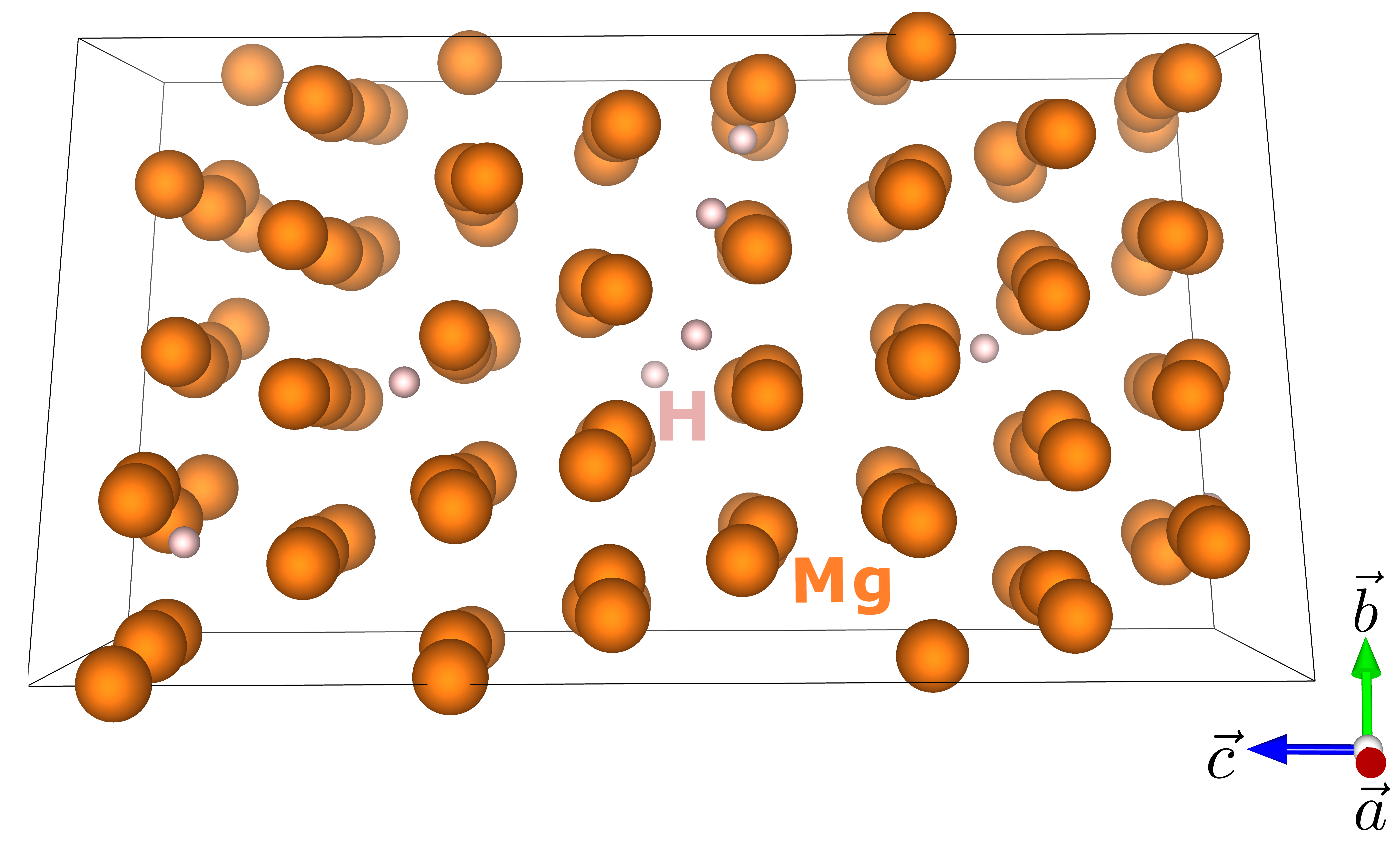}
\caption{Perspective view of a MLFF-MD frame, at 673~K, of the $4\times4\times4$ MgH$_{0.0625}$ supercell employed in the calculations. Mg atoms are displayed in orange and H atoms in pink.}
\label{poscar}
\end{figure}

The Density Functional Theory (DFT), MLFF-MD and \blue{climbing-image-NEB\cite{cineb} (ciNEB)} calculations were performed using VASP \blue{version 6.4.2 and version 6.4.3~\cite{Kresse1996, Kresse1999, mlff1, mlff2}, respectively}.
We utilized a $4\times4\times4$ supercell, shown in Fig.~\ref{poscar}, comprising of 128 Mg atoms in hcp crystal symmetry, with 8 H atoms \blue{(MgH$_{0.0625}$)} randomly distributed in the lattice.  
\blue{We performed non-spin-polarized calculations,} at the Perdew-Burke-Ernzerhof (PBE) functional level of theory~\cite{Perdew1996}, using an energy cutoff of 600 eV \blue{and a $4\times4\times4$ k-points grid}, with convergence threshold of 0.1 meV\blue{, employing a Gaussian smearing with a sigma value of 0.05~eV. To ensure consistency and avoid any discrepancies with our database, we adopted the same convergence parameters as those used in the Materials Project Database.  For VASP-MLFF we used the default parameters}.  
The temperature was incrementally raised from 0 to 700 K over a 0.2 ns interval with VASP on-the-fly MLFF-MD. 
In this setup, derived configurations—including structure energy, forces acting on each atom, atomic coordinates, stress tensor, and lattice parameters—were used to train the interatomic potential. Whenever the Bayesian error surpassed the set fixed threshold, VASP reverted to DFT to generate a new configuration in the database. Such on-the-fly procedure allows the model to use the accumulated ab-initio configurations to gradually improve the predictions on subsequent steps. 
The threshold value of 5~meV/Å  has been set to ensure a collection of diverse and well-spread representation of structures in the database from the explored configurational phase space.
During the thermalization phase, we opted for an NpT ensemble while constraining the cell shape, employing the Langevin thermostat with a friction coefficient for the lattice and atomic degrees-of-freedom equal to 10 ps$^{-1}$. 
We applied zero external pressure and a time step of 1~fs.
At key temperatures of 300~K, 480~K and 673~K, we conducted further 100~ps long NpT simulations in training mode to accumulate additional configurations. 
A total of over 3,700 ab-initio configurations have been stored, fewer than 1,000 being recorded at each constant temperature, and the remaining configurations captured during the ramping phase. 
Subsequently, we switched to MLFF-MD in run mode and determined the average lattice volume over a 100~ps period at fixed temperatures of 300~K, 480~K, and 670~K. The average volume configurations were then utilized to conduct NVT simulations. Following the 100~ps NVT simulation, we extracted \blue{three structures with an energy close to the average value of the NVT run. To avoid any  correlations, we ensured that the time interval between each of them was at least 10 ps.
Subsequently, starting from those, we performed three distinct} NVE simulations, where we computed the mean squared displacements (MSD) of hydrogen atoms as the ensamble average of
\begin{equation}
    \mathrm{MSD}(t)= \frac{1}{T-t} \int_0^{T-t}[\mathbf{r}(t+\Delta)- \mathbf{r}(\Delta)]^2 d \Delta. 
\end{equation}
Where $T$ is the total simulation time, and $\mathbf{r}$ is the trajectory of the atoms under analysis.
\blue{During the final NVE run a reduced time step of 0.5~fs was employed to reduce energy fluctuations}. 
All the MSD were constructed over NVE trajectories of 1~ns, and used to extract the diffusion coefficient $D$ by fitting the linear part of the function with the Einstein relation 
\begin{equation}
\mathrm{MSD}(t) = 6Dt
\end{equation}
\blue{by fitting the initial 0.2~ns linear region of the simulation. The value of $D$ at each temperature 
was computed as the average value of the three NVE replicas. The associated uncertainty was computed as the standard error of the mean SEM = SD/$\sqrt{3}$, where SD is the standard deviation, and 3 the
number of our collected predictions. An example of MSD fitting procedures is included in the Supplementary Figure SF1. \\
The ciNEB calculations were performed using a 2x2x2 supercell with one Hydrogen positioned along the main paths between interstitial sites identified in the literature~\cite{dftneb}. A visualization of those sites is reported in SF3. Also, every computation was performed using ten images and a 0.01~eV/\AA~forces convergence criterion.}

\subsection{Universal inter-atomic potentials UIPs}
\begin{figure}[t!]
    \centering
    \includegraphics[width=0.6\linewidth]{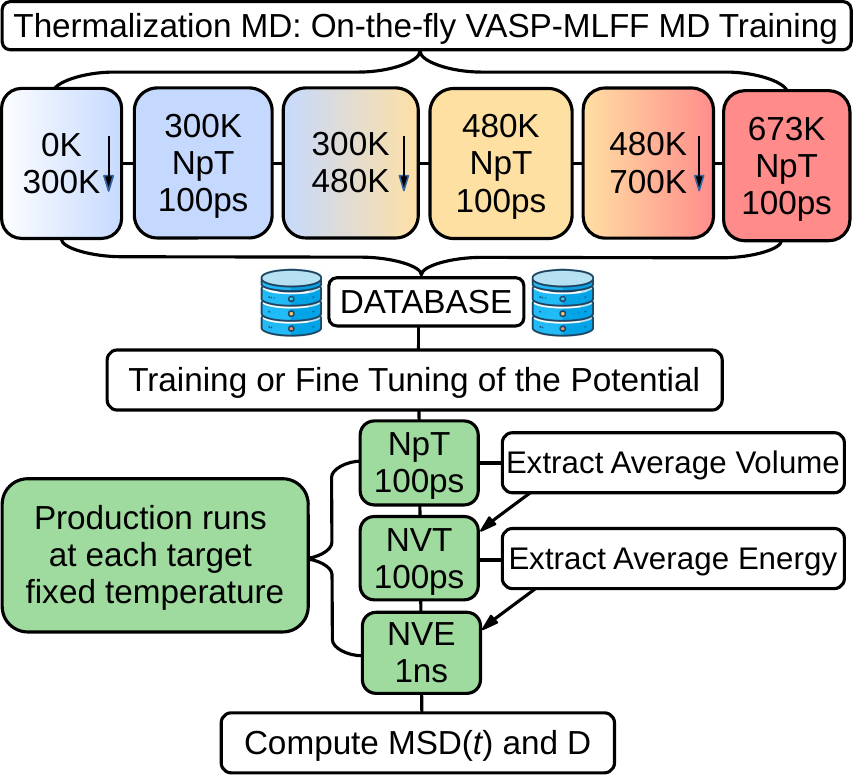}
    \caption{\textcolor{black}{Methodological protocol employed to improve the performance of interatomic potentials and obtain dynamical properties. The database of configurations is built both during NpT-MD thermalizations of the system from 0~K to 700~K via active learning of the VASP-MLFF, and at target temperatures (300~K, 480~K and 673~K). Subsequentially, the machine learned potentials are fine tuned or trained, and after system-equilibration the MSD and the diffusion coefficient D at fixed T are computed. 
    }}
    \label{workflow}
\end{figure}

We consider two state-of-the-art best performing~\cite{riebesell2023matbench} pre-trained UIPs: MACE~\cite{batatia2023foundation}, based on an equivariant message passing neural network, and CHGNet~\cite{chgnet}, a graph-based neural network. 
The models are imported in their pre-trained versions on the Materials Project (MP) relaxation trajectories database~\cite{jain2013commentary}, comprising 1.6 millions crystal structures with the associated energies, forces, stresses and magnetic moments.
In order to tackle the predicted dynamical properties from such UIPs, we followed the same 
\blue{NEB and}
MD procedure (excluding the on-the-fly training phase) as explained in the previous section. Subsequently, CHGNet and MACE were fine-tuned on the VASP-DFT generated data obtained during the on-the-fly MLFF-MD. 
MACE was also trained from scratch on the dataset. 
\blue{Both traning and fine-tuning of all the models were performed using the standard procedure and parameters suggested by the developers in the respective github repositories.
During CHGNet's MD simulations, we observed the need of a 0.5~fs time step to stabilize the dynamic also in the NPT and NVT runs, avoiding drifts in temperature for NVE. In this regard, we show examples of temperature oscillations during NVE simulations from 0.5~fs and 1.0~fs cases in the SF4 and SF5.}
To perform the MD simulations we employed LAMMPS~\cite{LAMMPS} and ASE~\cite{ase}, for MACE and CHGNet respectively.  \textcolor{black}{A schematic view of the employed workflow is represented in Fig.~(\ref{workflow}).}
\blue{Furthermore, using the best performing model, (MACE\_FT), we followed the same protocol to estimate $D$ at three additional hydrogen concentrations, containing 10~H (MgH$_{0.03125}$), 6~H (MgH$_{0.046875}$) and 4~H (MgH$_{0.046875}$), respectively.}
To assess the performance of every model, we evaluated the Root Mean Square Errors (RMSE) for the prediction of energy and forces over a test dataset of \blue{1200} configurations. \blue{These were randomly sampled, with proper equal spacing in time, from the NVE replicas of the MACE\_FT model, with 400 samples for each temperature. This validation process was independently repeated for each different hydrogen content to test the models' accuracy beyond the training concentration.
At last, NEB calculation on the same path studied using VASP were performed with every model using the ciNEB implementation of ASE employing the same number of images and convergence criteria.}

\section{Results and Discussions}
\begin{figure*}[h!]
    \centering
    \includegraphics[width=\linewidth]{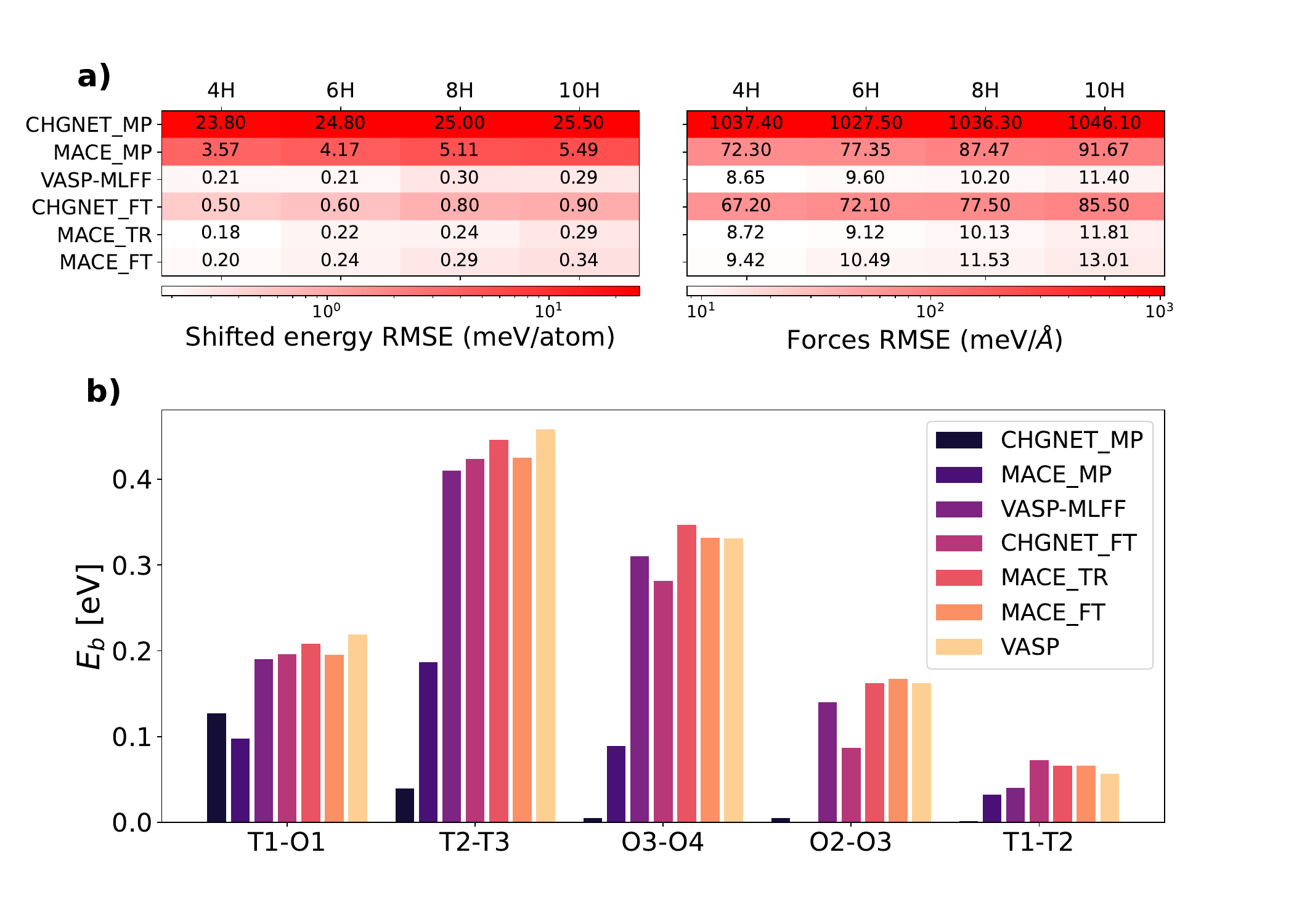}
    \caption{
    \textcolor{black}{a) RMSE values for the predictions of energies and forces on the validation set from every model considered in the study.
    \blue{These results are given at different concentrations, obtained using a supercell with a variable number of H atoms, removing the corresponding constant energy shift by setting the mean of all the configuration at same concentration to zero}. 
    The subscript 'MP' refers to the pre-trained model on the Material Project database, 'TR' to the one trained from scratch and 'FT' to the fine-tuned version.
    \blue{b) Energy barrier values associated to the hydrogen transition between different octahedral (O$_i$) and tetragonal (T$_i$) sites, obtained through ciNEB calculations, using different ML-potentials and DFT reference (in yellow).}
    }}
    \label{tab:perfmodels}
\end{figure*}

\begin{figure}[b!]
    \centering
\includegraphics[width=1\linewidth]{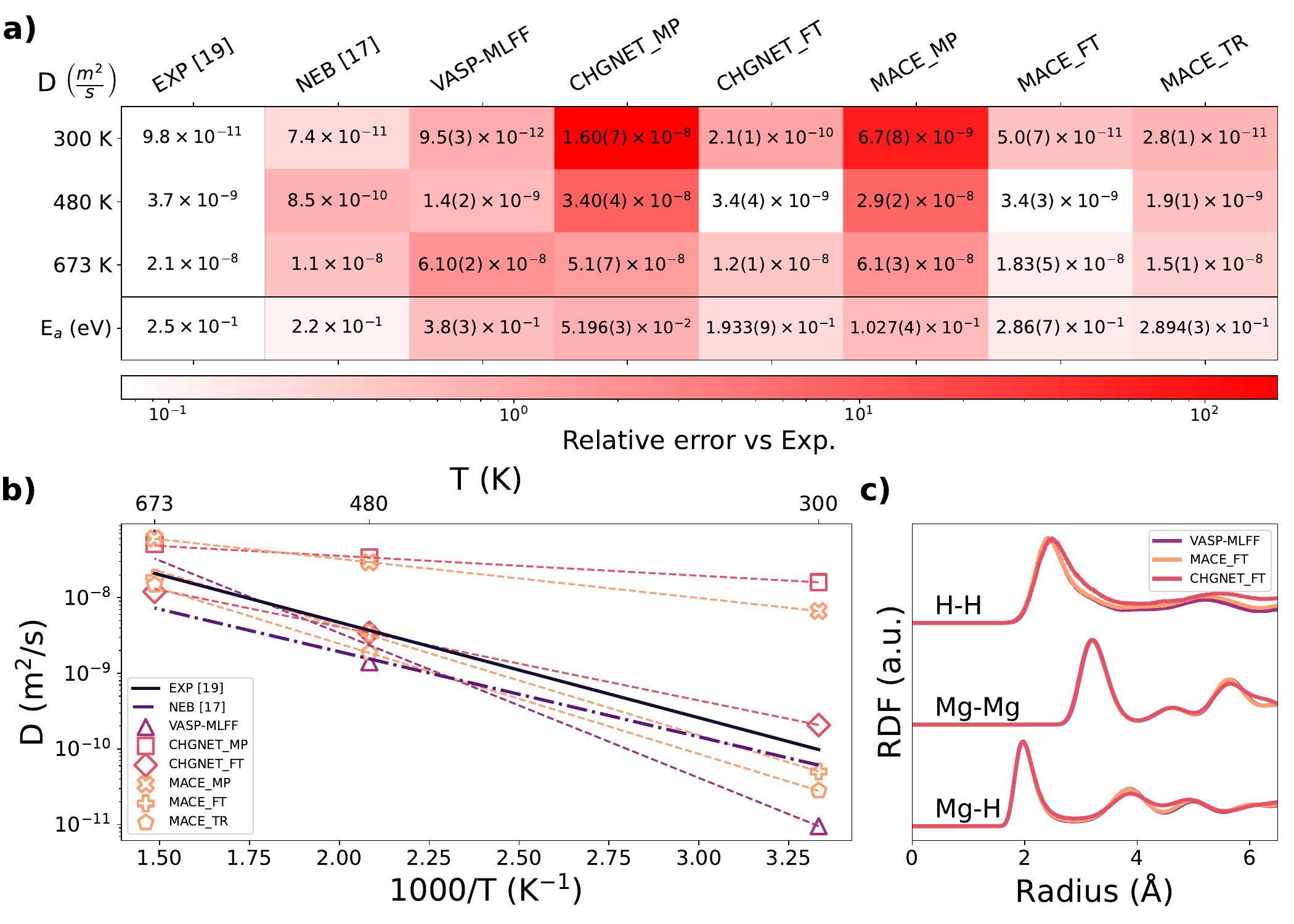}
    \caption{a) Diffusion coefficient values of hydrogen in MgH$_{0.0625}$ 
    at three different constant average temperatures of 380~K, 480~K and 673~K, and the corresponding activation energy, \blue{comparing} results achieved via different potentials in our investigations, and previous studies\blue{\cite{nishimure,dftneb}}. \blue{The uncertainty is reported in bracket for the last digit.} The colorbar highlights in logscale the relative deviation with respect to the experimental values. 
    b) Comparison between the dependence of the diffusion coefficient with temperature, for \blue{ML-models}, the NEB\cite{dftneb} and experimental results\cite{nishimure}. \blue{The experimental curve is extrapolated from data obtained at 474-493~K}.
    c) Radial distribution function of atoms pairs at 673~K during 1~ns long NVE simulations. VASP-MLFF and MACE show a remarkable agreement over the whole domain for every pair, while CHGNet starts to differ at larger distances.}
\label{fig:GrandeTabella}.
\end{figure}

\begin{figure}[h!]
    \centering    \includegraphics[width=0.73\linewidth]{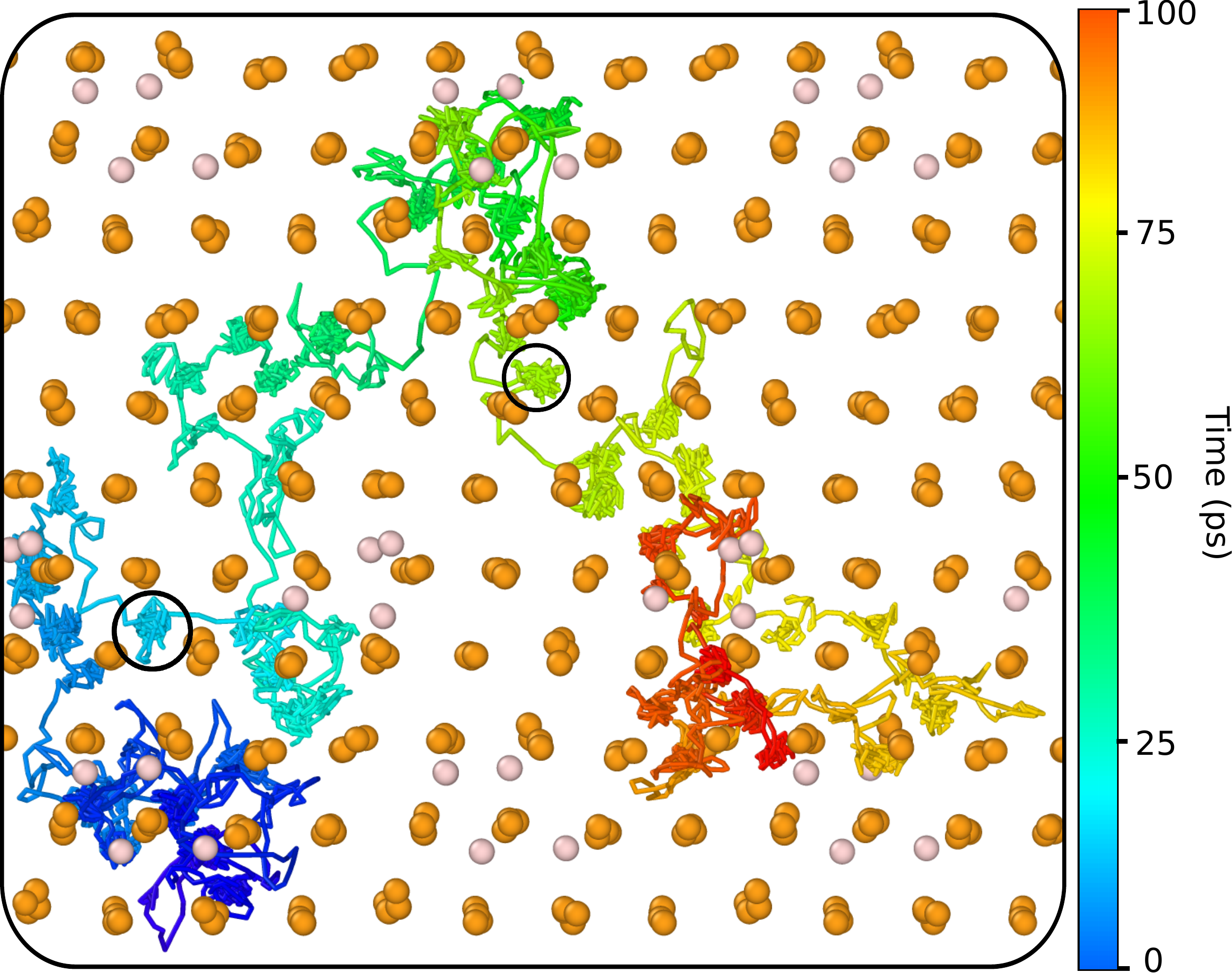}
    \caption{Diffusion path of a representative hydrogen atom in MgH$_{0.0625}$ during 100~ps at 673K, depicted using a color-gradient line to represent the progression of time, from blue to red. \textcolor{black}{The black circles highlight the interstitial regions where H atoms tend to oscillate around Mg lattice site before continuing their diffusing trajectory.}}
    \label{diff}
\end{figure}

\begin{figure}[h!]
    \centering    \includegraphics[width=\linewidth]{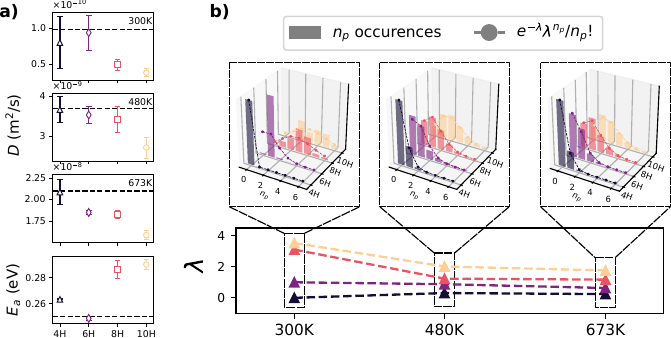}
    \caption{\blue{
    Results obtained from runs at different hydrogen content using MACE\_FT. a) Diffusion coefficients at different temperature and H concentration, on top, and obtained activation energy at different concentration, on the bottom. The target experimental value from~\cite{nishimure} is reported as the dashed line in every plot. b) Comparison of the simulated hydrogen pair distributions ($n_p$ occurrences, bars) with theoretical Poisson distributions (points and dashed lines) at three temperatures (300~K, 480~K, and 673~K) for varying hydrogen concentrations. The lower panel shows the $\lambda$ parameter (average number of pairs) as a function of temperature and concentration. 
}
}
    \label{hhpair}
\end{figure}

The results reported in Fig.~(\ref{tab:perfmodels}) show how the predictions of the VASP-MLFF \blue{achieve} an accuracy \blue{below} 0.3~meV/atom for energies and \blue{below} 10~meV/Å  for forces, respectively, as reasonably expected compared to other studies involving MLFF-MD~\cite{quantumverdi, quantumluigi,verdi2021thermal,zirconia}.
On the other hand, the UIPs pre-trained on the MP-database, respectively named on CHGNet\_MP and MACE\_MP , failed to reach such level of accuracy by more than one order of magnitude.
The discrepancy was significantly reduced after fine-tuning the two models on the 
VASP-generated
database: the error obtained shows that the CHGNet\_FT performance reaches a level comparable with the VASP-MLFF, and MACE\_FT even outperforms it. 
The performance differences between CHGNet\_FT 
and MACE\_FT 
may stem from the equivariant architecture employed by the latter. MACE turned out to be highly data-efficient, leading to better fine-tuning results over small datasets, compared to the more data-hungry architectures like CHGNet.
The MACE model was also trained from scratch on the VASP-DFT configurations (MACE\_TR), achieving slightly smaller errors on 
forces 
with respect to MACE\_FT.
\blue{
Please note that all energy values at concentrations different from the training concentration exhibit a systematic shift dependent on the hydrogen content. However, the underlying physics of hydrogen dynamics is unaffected, as this shift remains consistent throughout simulations with a constant number of particles and does not influence the forces. Consequently, the models are still capable of accurately describing the hydrogen dynamics. For completeness, we also present the pristine validation and the corresponding R$^2$ values, as shown in SF2.}\\
\blue{The ciNEB calculations conducted with the ML-potentials, allowed to estimate the transition energy barrier $E_b$ between various octahedral and tetragonal hydrogen interstitial sites, and 
to further validate the accuracy of the ML potentials. In fact, we compared the obtained $E_b$ values through these methods with the VASP-DFT benchmark results. 
Our findings, shown in Fig.     
\ref{tab:perfmodels}-b), reveal that the universal potentials perform significantly less accurately in predicting the energy barriers. However, the performance of the trained and fine-tuned counterpart dramatically improve, producing results that are much closer to the VASP-DFT reference values.
The VASP-MLFF method also gave excellent agreement with the DFT benchmark. As expected, the model with better accuracy on the validation set, especially on the forces, obtained results closer to the DFT values.
}
\\The diffusion coefficient \blue{$D$} predicted from every model during the NVE runs is reported in Fig.~\ref{fig:GrandeTabella}-a), and compared with experimental~\cite{nishimure} and NEB~\cite{dftneb} results, at each investigated temperature. 
\blue{The experimental data from Nishimura et. al. were collected in the temperature range of 474-493~K and then extrapolated at higher and lower temperatures using an Arrhenius relation.}
The results clearly show that our procedure provides multiple solutions with excellent experimental agreement, outperforming NEB computations~\cite{dftmd} that until now \blue{has} represented the standard for such applications.
This holds true for the MACE\_FT potential in particular, which not only predicts the correct order of magnitude across all temperatures, but significantly agrees at \blue{ 480~K}.
The VASP-MLFF instead, shows a \blue{good} agreement at 480~K and 673~K, while missing the room temperature by one order of magnitude. \blue{Concerning} CHGNet\_FT, it provides \blue{remarkable}  agreement with the experimental value at 480~K, but at the lowest 
temperatures 
underestimates the result by one order of magnitude. 
As expected from the error analysis on energies and forces, both the pre-trained versions of the UIPs result in much lower agreement, with deviations of at least one order of magnitude from experimental values at most temperatures.
A closer inspection of the temperature dependence of the diffusion coefficients, as shown in Fig.~\ref{fig:GrandeTabella}-b), reveals that CHGNet\_FT still outperforms the VASP-MLFF by better reproducing the Arrhenius curve observed in the experiments, while the MACE\_FT solutions outperform both.
This shows how with smaller errors, the deep networks are capable to better represent the shape of the energy landscape, with respect to the Bayesian alternative.
In this regard, further quantitative analysis can be provided by comparing the predicted value of the \blue{activation} energy \blue{E$_a$}, obtained from the linear fits in the Arrhenius plots. In particular, we achieved values of \blue{0.4~eV} for VASP-MLFF, \blue{0.19~eV} for CHGNet\_FT and \blue{0.28~eV} for MACE\_FT, where the experiment value corresponds to 0.25~eV \cite{nishimure}.
MACE\_FT clearly excels at representing the energy landscape of the system.\\
\blue{Furthermore, we computed $D$ at various hydrogen concentrations using MACE\_FT, along with the associated
E$_a$, as shown in Fig.~\ref{hhpair}-a). The results reveal a decrease in hydrogen mobility with increasing concentration, leading to a corresponding reduction in the diffusion coefficient. Simultaneously, the activation energy increases with higher hydrogen densities.
The evaluation of $D$ across different concentrations enabled a comparison with experimental values, demonstrating that the results for lower concentrations, specifically MgH$_{0.03}$ and MgH$_{0.045}$, show the best alignment. It should be noted that Nishimura et al.\cite{nishimure} did not report the hydrogen concentration within their samples. 
These findings suggest that the target hydrogen concentration in Nishimura's experiments may be lower than 0.0625, assumed in previous DFT studies\cite{dftneb}.}
\\Further analysis have been performed on the dynamics of the system by evaluating the radial distribution function (RDF) in all of the NVE run performed.
We report in Fig.~\ref{fig:GrandeTabella}-c) the predicted behaviour by the best performing models in the system at 673~K, while other temperature cases can be found in \blue{Fig. SF6}. 
A very good agreement between MACE\_FT and VASP-MLFF is found, while the RDF of CHGNet\_FT departs from the others at larger distances, by smoothing out peaks.
From such curves it is possible to retain information about the behaviour of hydrogen during the simulation.
Proceeding in order with increasing distance radius, the first peak appears just before 2 Å in the Mg-H curve, in correspondence of the average distance of one Magnesium atom from the center of the nearest octahedral sites on which hydrogen tend to sit~\cite{dftmd}.
The second peak, belonging to the H-H pair, is very pronounced at around 2.6~\AA, reflecting a correlation between hydrogen atoms at this distance.
This behaviour finds agreement in the literature for molecular dynamics with magnesium hydride nanoclusters \cite{cluster}.
In fact, the distance of 2.6~\AA~corresponds to the one between two octahedral sites along the c-direction~\cite{dftmd}, as also reported in literature~\cite{cluster}, implying how hydrogen tend to sit on near sites.
To highlight such behaviour we also report in Fig.~(\ref{diff}) the extensive diffusion path of a representative hydrogen atom within the magnesium matrix during a 100~ps simulation, at 673~K. The trajectory is unwrapped in the replicas of the periodic images to enhance visibility and interpretation. 
The color gradient serves as a temporal marker, with blue indicating the initial position of the hydrogen atom at the beginning of the simulation (t = 0 ps) and red indicating its position at the end of the studied interval (t = 100 ps).
Intermediate colors (cyan, green, yellow, and orange) represent the progression of time between these two extremes, providing a visual cue for the temporal evolution of the atom's diffusion path.
The black circles highlight interstitial regions where the hydrogen atom tends to oscillate around the magnesium sites, indicating temporary trapping sites within the lattice structure, before continuing its diffusion trajectory.
The third significant peak in the RDF is observed around 3 Å in the Mg-Mg curve, consistently with the typical magnesium distances in hcp structures. Multiple smaller peaks indicate further neighbor interactions in the crystal lattice.
Analogous results were found in the RDF at 300~K and 480~K, where the peaks are
sharper due to the reduced effect of thermal motion, see SF6. In particular, higher pronounced RDF at lower temperature indicates that hydrogen tend to spend more time in the vicinity of Magnesium, and diffuse less in the crystal structure.
\\
\blue{The analysis of hydrogen pair ($n_p$) distributions presented in Fig.~\ref{hhpair}-b) provides a clear insight into the dependence of pair formation on temperature and concentration. At elevated temperatures, such as 480~K and 673~K, the simulated distributions align closely with the theoretical Poisson distribution, calculated as:
$$
f(n_p; \lambda) = \frac{e^{-\lambda} \lambda^{n_p}}{n_p!},
$$
where $\lambda$ represents the average number of pairs. This agreement demonstrates the random nature of hydrogen pairing, indicating a lack of correlation and independence from the initial atomic configuration. Furthermore, the convergence of $\lambda$ values for different concentrations as temperature increases underscores this intrinsic randomness. 
Conversely, at 300~K, a deviation from the Poissonian behavior emerges, with $\lambda$ showing a stronger dependence on hydrogen concentration. This behavior can be attributed to the reduced mobility of hydrogen atoms at low temperatures, where the thermal energy is insufficient to overcome the energy barriers for interstitial site transitions. Consequently, the hydrogen atoms remain in proximity to their initial positions, leading to non-equilibrated pair distributions within the simulation timescale. These results suggest that, while the system approaches a random distribution at higher temperatures, longer simulations would be required to achieve equilibration and Poissonian behavior at room temperature for low-mobility regimes.}

\section{Conclusion}
Our investigation enabled to thoroughly characterize the kinetic properties and mobility of hydrogen within a structured environment, such as pure magnesium, across various temperatures 
through a rigorous and efficient methodology. 
\textcolor{black}{We performed a systematic and comparative study of ML-based interatomic potential MD schemes, particularly focusing on Bayesian versus equivariant and graph neural networks (MACE and CHGNet), under different training modes (universal and fine-tuned).}
The results were validated by estimates of the diffusion coefficient 
\blue{and the activation energy} 
values, which showed excellent agreement with experimental data.
The obtained results proved that the VASP-DFT configurations, collected during the MLFF on-the-fly training, represent a complete set for the accurate modeling of inter-atomic interactions, between hydrogen and magnesium atoms. 
This strategy could be consistently applied to generate a comprehensive dataset for the proper training of existing or forthcoming potentials. 
In fact, we identified the limitations of pretrained UIPs in studying materials with diffusing defects, due to the absence of representative high temperatures, defective and metastable states in their datasets. 
Particularly, we demonstrated the ability of state-of-the-art machine learning models to achieve DFT-level accuracy, after fine-tuning on actively learned DFT-configurations, highlighting the importance of focusing on efficient dataset-building methods and their quality. 
Specifically, the importance of including defected and transition-state configurations, as well as the role of transfer-learning, allowing pre-trained solutions to adapt to new systems. 
\blue{The obtained data offer valuable new insights into the collective dynamical properties at varying hydrogen concentrations, accurately predicting a decrease in hydrogen mobility as concentration increases. Additionally, the statistical analysis of hydrogen pair distributions indicates the system approaches a random distribution at higher temperatures. However, longer simulations would be necessary to achieve equilibration and Poissonian behavior at room temperature in low-mobility regimes.}

These progresses not only improve our understanding of hydrogen interactions in magnesium, but also pave the way for future research into a broader range of multi-component systems and defected compositions. 
This is especially significant for systems with complex potential energy surfaces, where traditional ab-initio methods become impractical.
Successfully modeling the hydrogen diffusion mechanism in magnesium via machine learning-accelerated molecular dynamics could facilitate the study and discovery of new, more efficient materials for hydrogen storage and beyond it, contributing to the transition towards a greener and more sustainable energy future.

\newpage

\section*{Acknowledgments}
C. F. and A. A. acknowledges the ''Doctoral College Advanced Functional Materials – Hierarchical Design of Hybrid Systems DOC 85 doc.funds'' funded by the Austrian Science Fund (FWF) and by the Vienna Doctoral School in Physics (VDSP). For Open Access purposes, the author has applied a CC BY public copyright license to any author accepted manuscript version arising from this submission.
D. M. and S. P. were supported by the European Union Horizon 2020 research and innovation program under Grant Agreement No. 857470, from the European Regional Development Fund under the program of the Foundation for Polish Science International Research Agenda PLUS, grant No. MAB PLUS/2018/8, and the initiative of the Ministry of Science and Higher Education 'Support for the activities of Centers of Excellence established in Poland under the Horizon 2020 program' under agreement No. MEiN/2023/DIR/3795.
L.P. and C.F. acknowledges the National Recovery and Resilience Plan (NRRP),  Mission 4 Component 2 Investment 1.3 - Project NEST (Network 4 Energy Sustainable Transition) of Ministero dell’Università e della Ricerca (MUR), funded by the European Union – NextGenerationEU.
L.L. and C.F. acknowledges the NRRP, CN-HPC grant no. (CUP) J33C22001170001, SPOKE 7, of MUR, funded by the European Union – NextGenerationEU.
The computational results were obtained using the Vienna Scientific Cluster (VSC) and the LEONARDO cluster.
We acknowledge access to LEONARDO at CINECA, Italy, via an AURELEO (Austrian Users at LEONARDO supercomputer) project

\section*{Supplementary material}
Supplementary information and figures are available at website...\\
\blue{
The training dataset  
can be openly accessed at \href{https://figshare.com/articles/dataset/_b_ML_AB_b_/27922470?file=50852211}{https://figshare.com/articles/dataset/\_b\_ML\_AB\_b\_/27922470?file=50852211}.
}  

\section*{Author Contribution}
Conceptualisation: C.F., A. A., D. M. and L. L.
Methodology and calculations: A. A., L.L. and D. M.
Supervision: C.F.
Writing: A.A , D.M. , L. L. and C. F.
Discussion and reviewing: All authors

\section*{Competing Interests}
The authors declare no competing interests.

\newpage

\bibliographystyle{elsarticle-num}
\bibliography{./bib}

\end{document}


\title{{\it Supplementary Materials for}\\"Hydrogen Diffusion in Magnesium Using Machine Learning Potentials: a comparative study"}

\author{Andrea Angeletti}
\affiliation{University of Vienna, Vienna Doctoral School in Physics, Boltzmanngasse 5, 1090 Vienna, Austria}

\author{Luca Leoni}
\affiliation{Department of Physics and Astronomy 'Augusto Righi', Alma Mater Studiorum - Universit\`{a} di Bologna, Bologna, 40127 Italy}

\author{Dario Massa}
\affiliation{NOMATEN Centre of Excellence, National Center for Nuclear Research, 05-400 Otwock, Poland}

\author{Luca Pasquini}
\affiliation{Department of Physics and Astronomy 'Augusto Righi', Alma Mater Studiorum - Universit\`{a} di Bologna, Bologna, 40127 Italy}

\author{Stefanos Papanikolaou}
\affiliation{NOMATEN Centre of Excellence, National Center for Nuclear Research, 05-400 Otwock, Poland}

\author{Cesare Franchini}
\affiliation{Department of Physics and Astronomy 'Augusto Righi', Alma Mater Studiorum - Universit\`{a} di Bologna, Bologna, 40127 Italy}
\affiliation{University of Vienna, Faculty of Physics and Center for Computational Materials Science, Kolingasse 14-16, 1090 Vienna, Austria}
\maketitle

\noindent \blue{In this section we show supplementary figures to the main work:}\vspace{0.25cm}

\noindent \blue{Fig. SF1 shows the mean square displacement of one the VASP-MLFF replicas at the highest investigated temperature, and the component decomposition along three directions. Overall across the various simulations we have not observed a consistent difference in the behavior between the different components.}\vspace{0.25cm}

\noindent \blue{Fig. SF2 shows the un-shifted energy prediction errors of the models during their validation and the corresponding R$^2$ values.\vspace{0.25cm}}

\noindent \blue{Fig. SF3 shows the octahedral O$_i$ and tetragonal T$_i$ interstitial hydrogen sites in the magnesium matrix, between which we computed the associated energy barrier via ciNEB, with the ML-models and VASP-DFT.\vspace{0.25cm}}

\noindent \blue{Fig. SF4 shows the drifting in energy for the CHGNet\_MP model using a time step of 1~fs.
By reducing it to 0.5~fs the drifting disappears as can be see in Fig. SF5. 
Moreover, we can also observe the temperature fluctuations during the molecular dynamics simulations of CHGNet\_MP model, which turned to out to be most unstable model overall.\vspace{0.25cm}}

\noindent \blue{Fig. SF6 shows the Radial Distribution Function of all atom pairs for three ML-model at 480~K and 300~K.\vspace{0.25cm}}

\noindent \blue{Fig. SF7 shows estimates of the MSD at different temperature for the MACE\_FT model using different time lags (parameter $\delta$ of MSD formula in Ref.~\cite{Pranami2015}). 
In particular, the results are calculated for the standard $\sigma = 1$ and the fully uncorrelated $\sigma = \tau$ cases, estimating also the diffusion coefficient to show a possible dependence on the parameter.
The two estimates showed nearly the same linear trend with differences mainly related to the lower statistic available to the $\delta = \tau$ case that makes it more noisy.
This is well reflected by the estimates of $D$ fitted in the range described previously for the $\delta = 1$ case and in a smaller range of 0 and 0.1~ns for the uncorrelated case, where the smaller range is needed due to the much higher noise in longer times.
The estimates deviates of at most 2.5~\% between the two cases still remaining perfectly compatible within the estimated error, leading to variations in the estimates of the energy barrier of less than 1~\%.
This shows how no major correlation is present inside the MSD of the case under study, making the values obtained with $\delta = 1$, corresponding to the formula reported in the main text, more reliable due to their much larger statistic that allows for a converged estimates in the simulation time under investigation.}

\begin{figure}[h!]
    \centering
    \includegraphics[width=\linewidth]{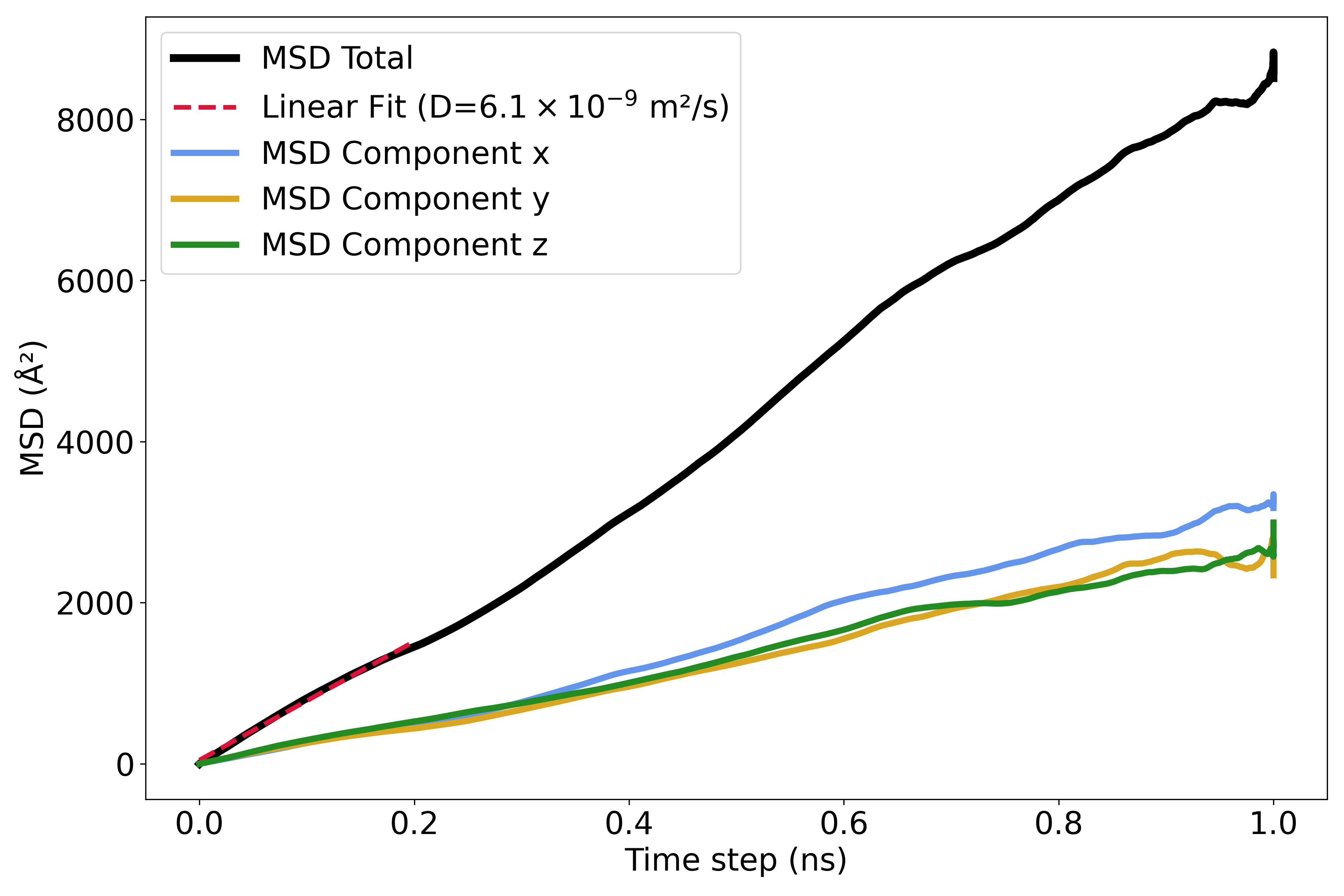}
    \caption{Example of MSD fitting procedure for the case of VASP MLFF model at 673K and the values along different directions.}
    \label{fig:enter-label}
\end{figure}
\begin{figure}[h!]
    \centering
    \includegraphics[width=\linewidth]{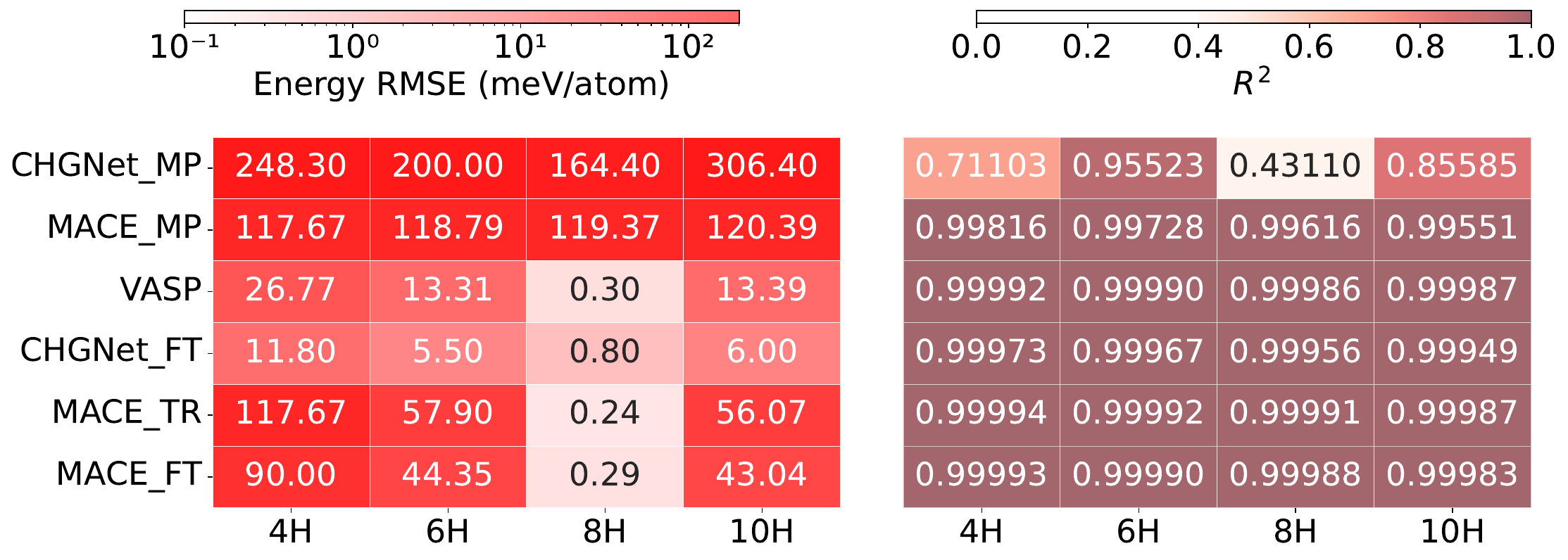}
    \caption{\blue{The left table shows the RSME for energies and forces for all models sand concentrations, without energy shifting. The corresponding R$^2$ values for the predictions are on the right table}.}
    \label{fig:enter-label}
\end{figure}
\begin{figure}[h!]
    \centering
    \includegraphics[width=0.7\textwidth]{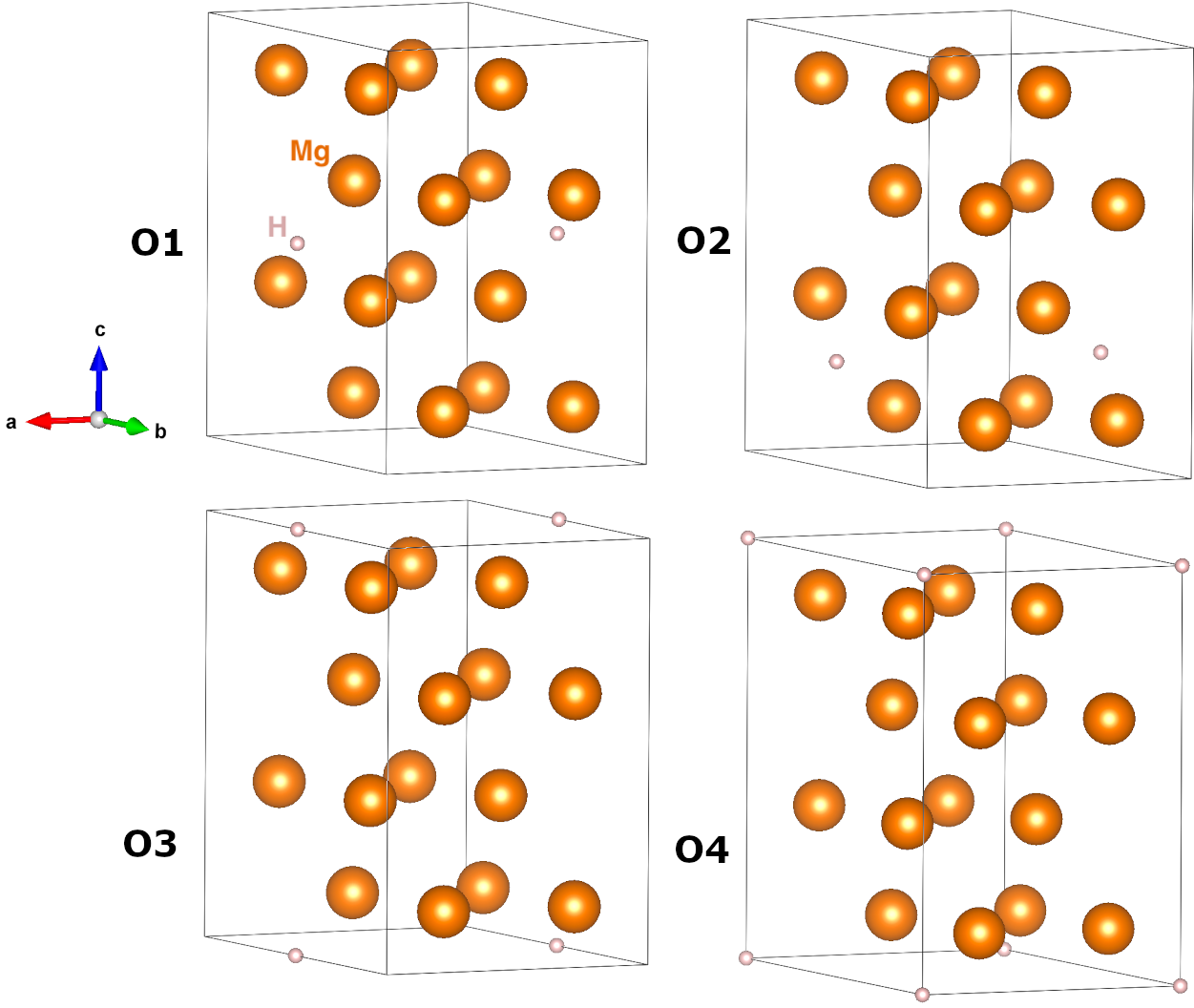}
    \includegraphics[width=0.7\textwidth]{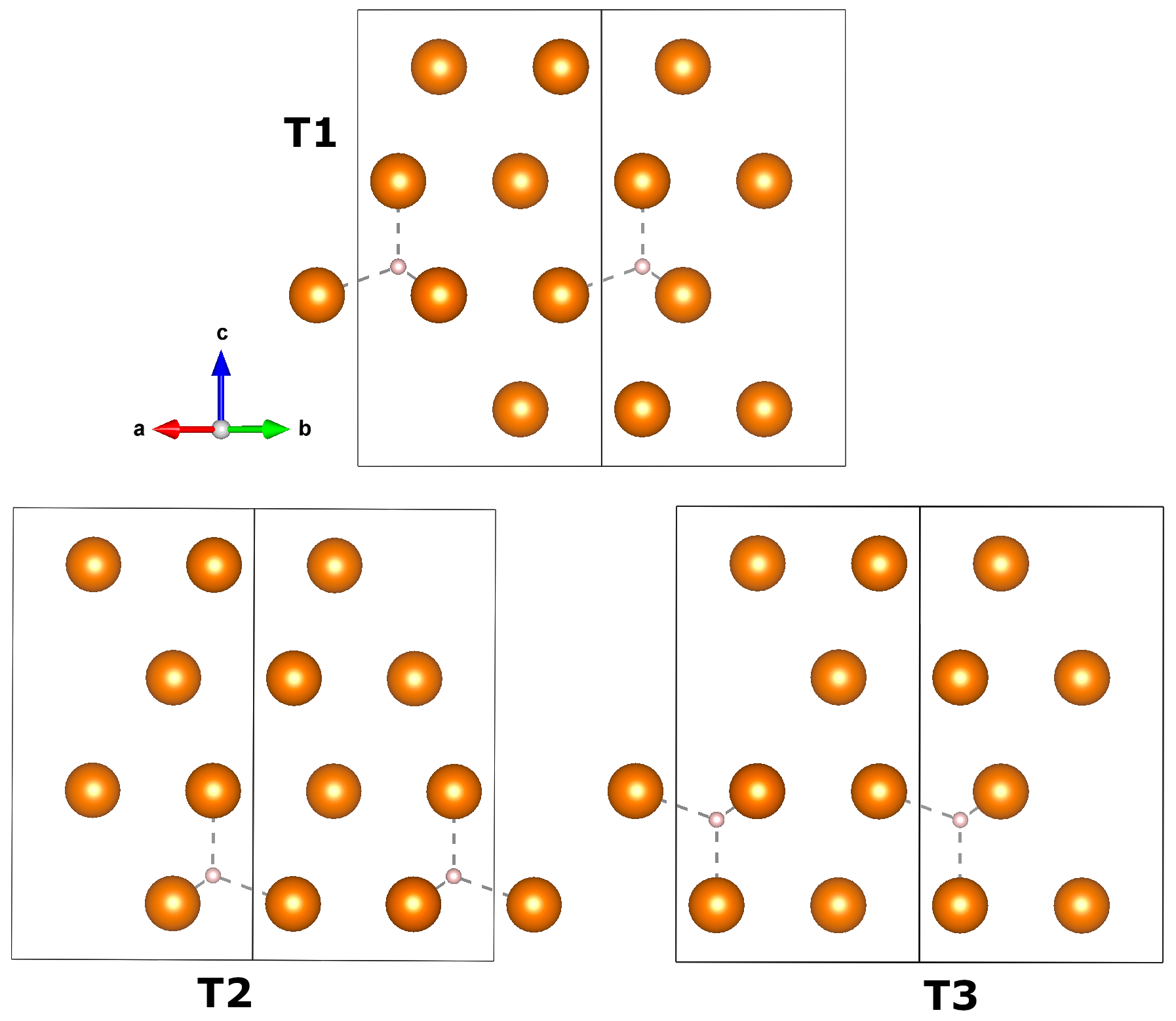}
    \caption{\blue{Interstitial states considered for NEB calculations with all models.}}
    \label{fig:NEBstates}
\end{figure}
\begin{figure}[h!]
    \centering
    \includegraphics[width=0.7\textwidth]{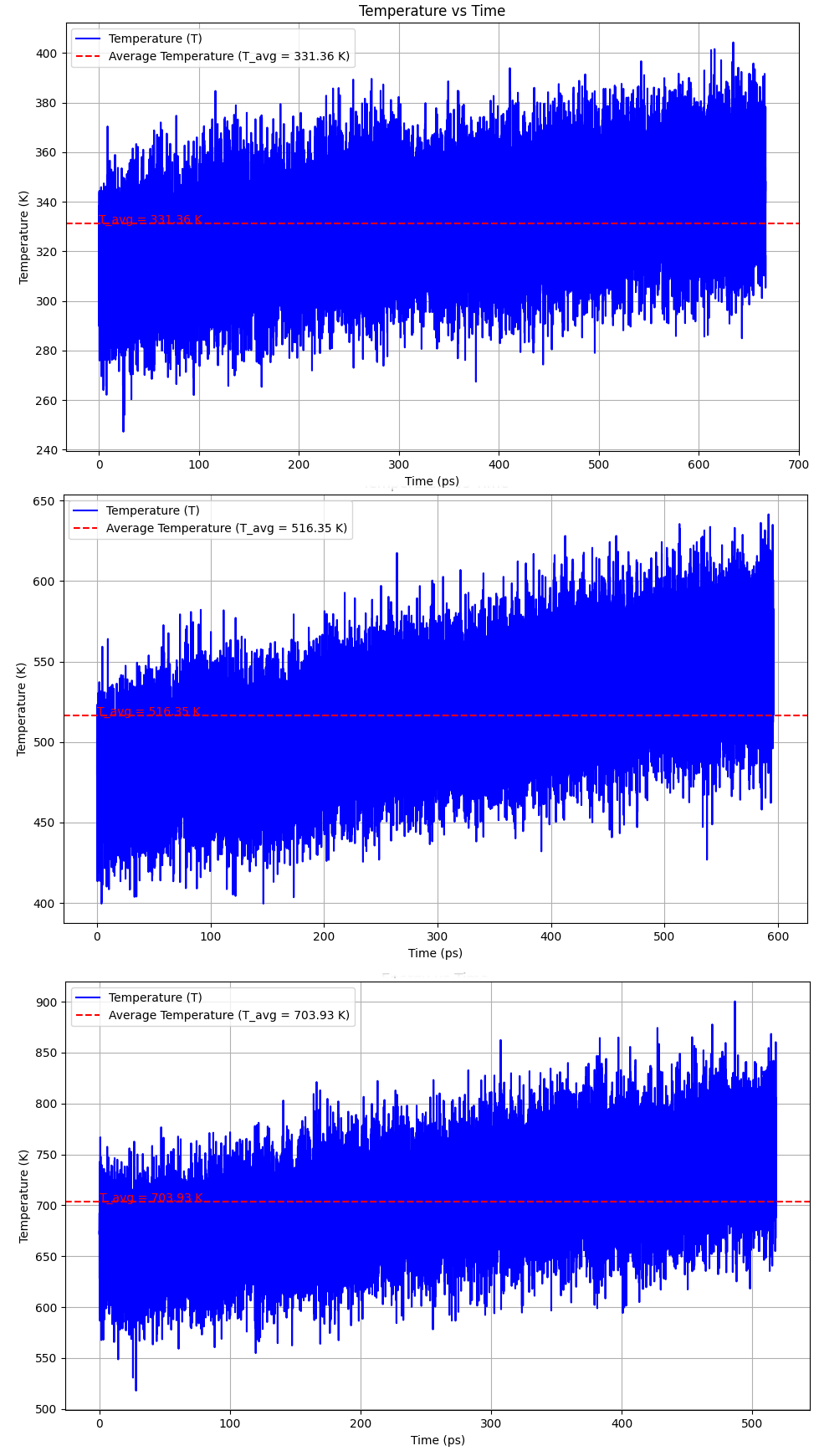}
    \caption{Example of drifting in the NVE temperatures for the case of 1~fs timestep simulations with the pretrained CHGNet\_MP model.}
    \label{fig:enter-label}
\end{figure}
\begin{figure}[h!]
    \centering
    \includegraphics[width=0.6\textwidth]{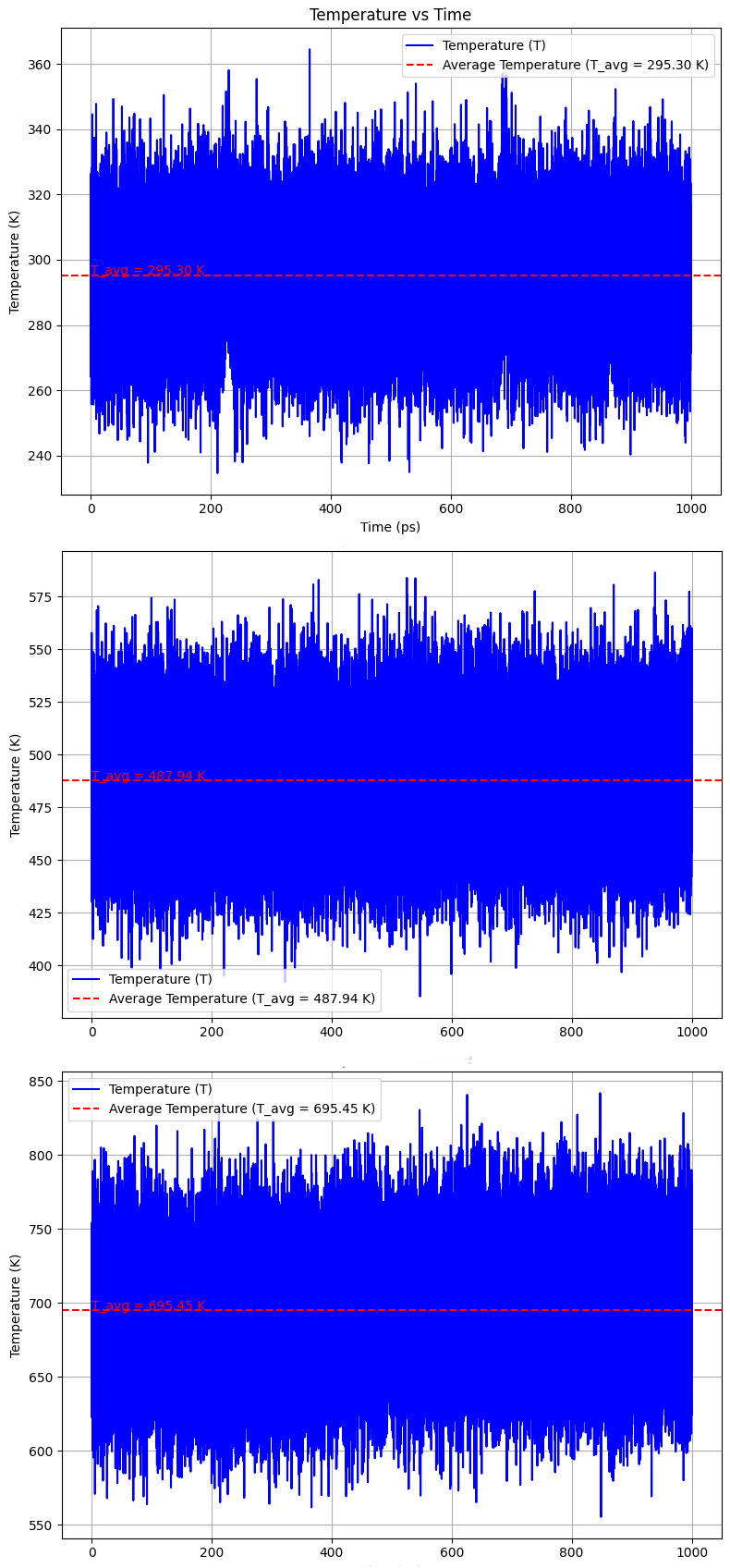}
    \caption{Example of stable temperature oscillations in the NVE simulations for the case of 0.5~fs timestep with the CHGNet\_MP model.}
    \label{fig:enter-label}
\end{figure}
\begin{figure}[h!]
    \centering
    \includegraphics[width=0.85\textwidth]{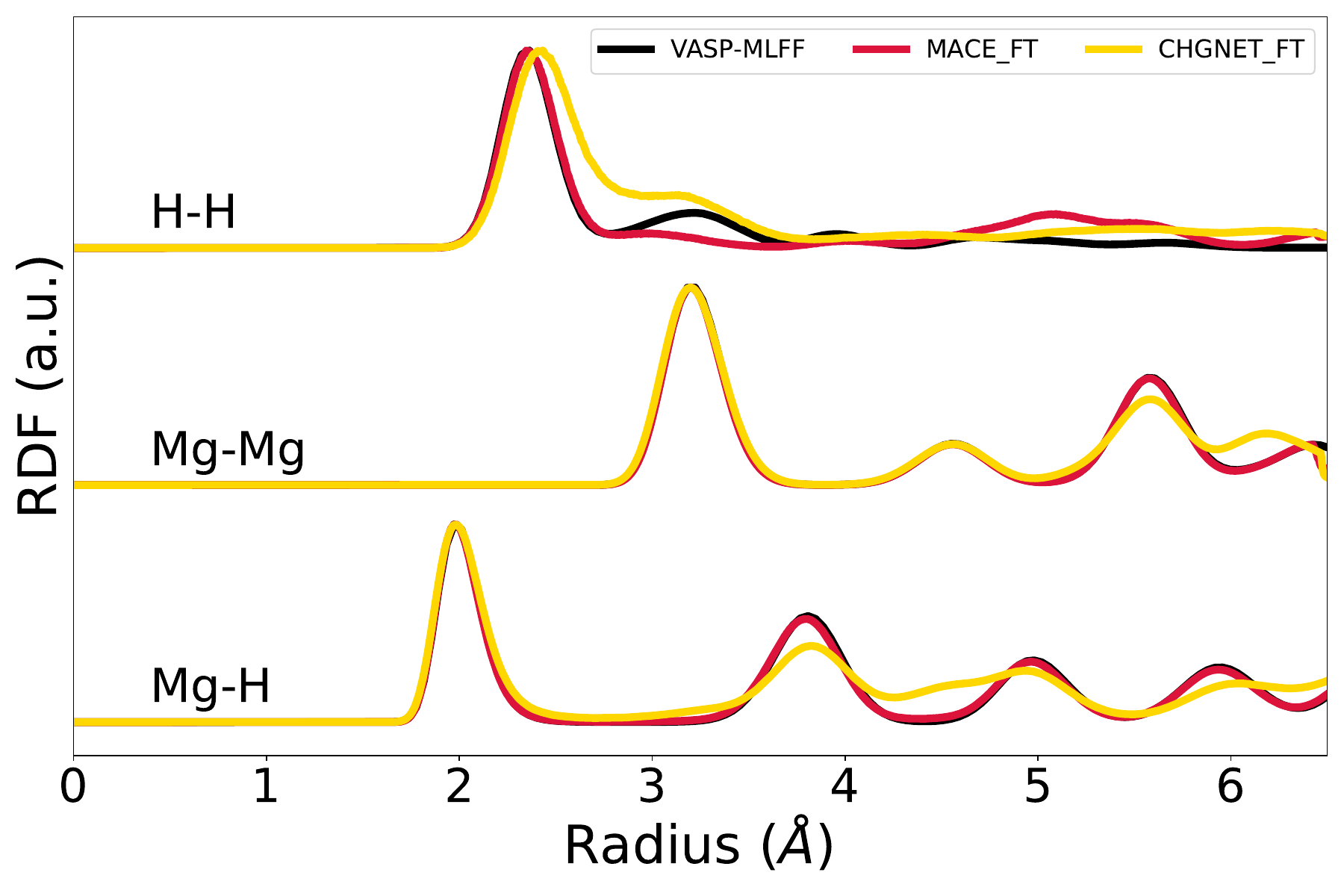}
    \includegraphics[width=0.85\textwidth]{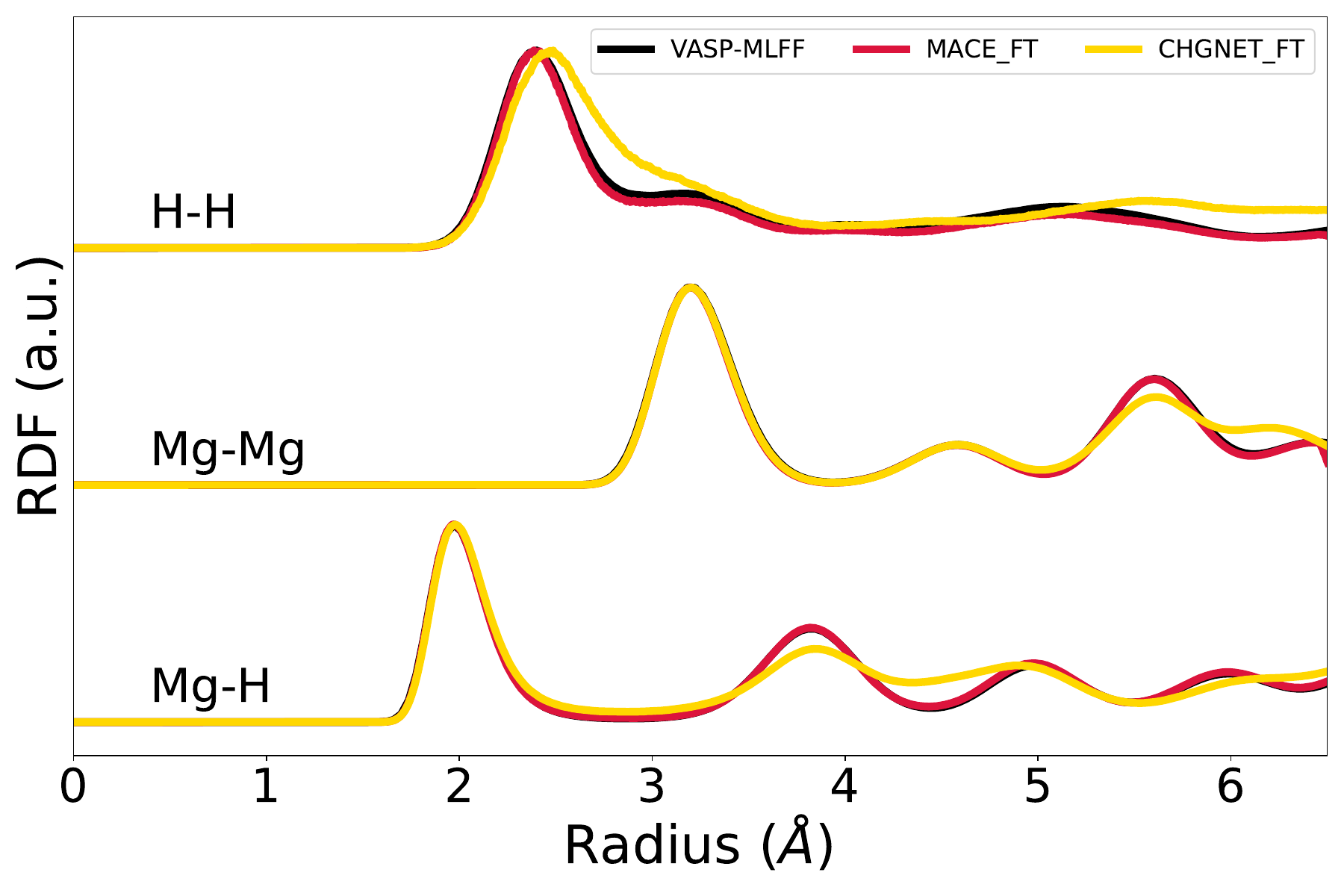}
    \caption{Radial distribution function at different temperatures for the best performing models, at 300~K on top and 480K at the bottom.}
    \label{fig:enter-label}
\end{figure}
\begin{figure}[h!]
    \centering
    \includegraphics[width=\linewidth]{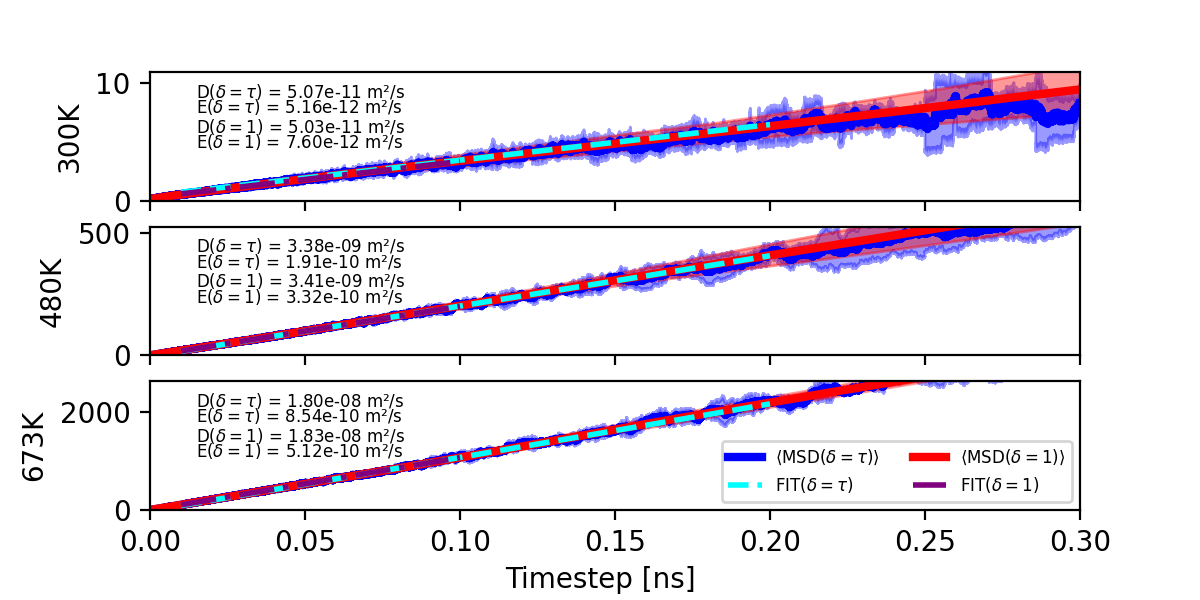}
    \caption{
    \blue{Values of the MSD at different temperatures, with following linear fit, for the MACE\_FT model estimated for different time lags, $\delta$, following the formula in Ref.~\cite{Pranami2015}. }
    }
    \label{fig:enter-label}
\end{figure}

\newpage
\bibliography{manuscript/bib}